\documentclass[%
nofootinbib,
 amsmath,amssymb,
 aps,
 prl,
 twocolumn,
 superscriptaddress
]{revtex4-1}

\usepackage{dcolumn}

\allowdisplaybreaks

\usepackage[dvipdfmx]{graphicx}
\usepackage{latexsym}
\usepackage{amsfonts}
\usepackage{amssymb}
\usepackage{amsmath}
\usepackage{bm}
\usepackage{mathrsfs}
\usepackage[dvipdfm]{hyperref}

\usepackage{slashed}
\usepackage{color}

\newcommand{\bB}{{\bm{B}}}

\newcommand{\tr}{ {\rm Tr} }
\newcommand{\Lag}{ {\mathscr{L}} }

\newcommand{\M}{ {\mathcal M} }

\newcommand{\ext}{ {\rm ext} }
\newcommand{\vac}{ {\rm vac} }

\newcommand{\ope}{ {\rm OPE} }

\newcommand{\pv}{{_{\scriptscriptstyle  \rm PV}}}
\newcommand{\ps}{ {\scriptscriptstyle \rm P} }
\newcommand{\V}{ {\scriptscriptstyle \rm V} }
\newcommand{\gam}{ {\scriptscriptstyle \gamma} }

\newcommand{\etac}{ {\eta_c} }
\newcommand{\Jp}{ {J/\psi} }

\newcommand{\ph}{ {\rm ph} }
\newcommand{\dir}{ {\rm dir} }

\newcommand{\id}{\mbox{1}\hspace{-0.25em}\mbox{l}}

\newcommand{\beq}{\begin{eqnarray}}
\newcommand{\eeq}{\end{eqnarray}}

\newcommand{\Whittaker}[3]{G\left( #1, #2, #3 \right)}

\def\simge{\mathrel{%
   \rlap{\raise 0.511ex \hbox{$>$}}{\lower 0.511ex \hbox{$\sim$}}}}
\def\simle{\mathrel{
   \rlap{\raise 0.511ex \hbox{$<$}}{\lower 0.511ex \hbox{$\sim$}}}}
\def\bigs{\mathrel{
   \rlap{\raise 0.531ex \hbox{$>$}}{\lower 0.531ex \hbox{$<$}}}}

\usepackage[normalem]{ulem}  

\renewcommand\sout{\bgroup \color{red} \ULdepth=-.5ex \ULset}

\begin{document}

\vspace*{-10mm}
\begin{flushright}
KEK-TH-1746, RIKEN-QHP-158
\end{flushright}
\vspace{-5mm}

\vspace*{-10mm}
\begin{flushright}
\end{flushright}
\vspace{-5mm}


\author{Sungtae Cho} \email{\tt sungtae.cho@yonsei.ac.kr}

\affiliation{
Institute of Physics and Applied Physics, 
Yonsei University, Seoul 120-749, Korea
}


\author{Koichi Hattori} \email{\tt koichi.hattori@riken.jp}

\affiliation{
Institute of Physics and Applied Physics, 
Yonsei University, Seoul 120-749, Korea
}
\affiliation{
Theoretical Research Division, Nishina Center, RIKEN, Wako, Saitama 351-0198, Japan
}
\affiliation{
RIKEN BNL Research Center, Bldg. 510A, Brookhaven National Laboratory, Upton, NY 11973, USA
}


\author{Su Houng Lee} \email{\tt suhoung@yonsei.ac.kr}

\affiliation{
Institute of Physics and Applied Physics, 
Yonsei University, Seoul 120-749, Korea
}


\author{Kenji Morita} \email{\tt morita@fias.uni-frankfurt.de}

\affiliation{Frankfurt Institute for Advanced
Studies, Ruth-Moufang-Str. 1, D-60438 Frankfurt am Main, Germany}
\affiliation{
Institute of Theoretical Physics, University of Wroclaw, 
PL-50204 Wroclaw, Poland
}
\affiliation{
Yukawa Institute for Theoretical Physics, Kyoto University, Kyoto 606-8502, Japan
}

\author{Sho Ozaki} \email{\tt sho@post.kek.jp}
\affiliation{
Institute of Physics and Applied Physics, 
Yonsei University, Seoul 120-749, Korea
}
\affiliation{
Theory Center, IPNS, High energy accelerator research organization (KEK), 
1-1 Oho, Tsukuba, Ibaraki 305-0801, Japan
}


\vspace*{5mm}
\title{QCD Sum Rules for Magnetically Induced Mixing between $\eta_c$ and $J/\psi$}


\date{\today}

\begin{abstract}
We investigate the properties of charmonia in strong magnetic fields by using QCD sum rules. 
We show how to implement the mixing effects between $\eta_c$ and $J/\psi$ 
on the basis of field-theoretical approaches, 
and then show that the sum rules are saturated by the mixing effects 
with phenomenologically determined parameters. 
Consequently, we find that the mixing effects are the dominant contribution 
to the mass shifts of the static charmonia in strong magnetic fields. 
\end{abstract}





\maketitle

Ever since the suppression of $J/\psi$ yields due to the color screening effect was
proposed as a signature of the formation of the quark-gluon plasma in
ultrarelativistic heavy-ion collisions \cite{MS,Hashi}, much
attention has been paid to the properties 
of heavy quarkonia under extreme environments. 
Meanwhile, extremely strong electric and magnetic fields
induced by 
injection of heavy-ions have been discussed recently \cite{KMW,Bestimates,Itakura_PIF},
because they could be 
new ingredients affecting experimental observables at the Relativistic Heavy Ion
Collider and the Large Hadron Collider.
A renewed interest arises in the
study of the heavy-quark (HQ) systems 
and 
their spectral densities 
in the strong fields \cite{MT,AS} as some of these states will likely form in an earlier time
after the impact
\cite{formation} where the fields still persist with 
large strengths.

The QCD sum rule (QCDSR) has been extensively used for investigating
the spectral density of the hadrons on the basis of the
fundamental quark and gluon degrees of freedom
\cite{SVZ792,RRYrev,Narison}. Remarkably, the QCD sum rules for
the HQ systems predicted the small mass splitting between $\eta_c$
and $J/\psi$ of the order of 100 MeV prior to the experimental 
confirmation of the $\eta_c$ mass \cite{etac,SVZ792,RRY81}.
While the properties of charmonium systems  in the vacuum are well
described by the Cornell potential model \cite{Cornell},
the advantage of using the QCDSR is that effects of external
environments on the correlation functions can be easily taken into
account from the modifications in the operator product expansion
(OPE) through moderate changes in values of quark and gluon condensates.
Moreover, for HQ systems, the modification involves only the
dimension-4 operators that are related to the energy momentum
tensor whose matrix elements are well estimated both at finite
temperature from lattice QCD \cite{ML,ML08,ML10,GMO,SGMO} and at
normal nuclear matter density from measurements in deep inelastic
scatterings \cite{KKLW}. Recently, it has been shown that even the
temperature dependence of the gauge invariant strength of the
charmonium wave function at the origin obtained from the QCDSR
supports that from solving the Schr\"odinger equation with a finite 
temperature free energy potential from lattice QCD
\cite{Lee:2013dca}.

In this Letter, we apply the QCDSR to investigate the mass spectra
of the lowest-lying bound states coupled to pseudoscalar (PS)
and vector HQ currents in external magnetic fields ($B$-fields);
that is, the $\eta_c$ and $J/\psi $ at rest. 
We put a special emphasis on how to take into account 
mixing effects in the spectral density, the so-called phenomenological side, and show how to distinguish
nonperturbative mass modifications from hadronic mixing effects
between $\etac$ and $\Jp$. Since mixing effects naturally arise in
external environments, as have been known for a
long time in various systems such as hydrogen atoms and
positronium in external electromagnetic fields, our results can be
generalized to various systems accompanied by 
mixing effects.
We note that our treatment of the mixing effects should be applied to the very recent QCDSR analysis
on $B$ mesons in strong $B$-fields \cite{Machado},
since the $B$ mesons are mixed with $B^\ast$ mesons.  
Our work demonstrates how to implement mixing effects in the QCDSR method, 
in particular for the HQ systems where both the
OPE and the phenomenological side are well under control, 
and thus provides a general guideline to include mixing effects in approaches based on correlation functions. 


We first begin by looking at the general results of the mixing
effect using effective Lagrangians. A three-point
vertex which can describe a radiative decay mode, $J/\psi \to
\eta_c + \gamma$, induces mass shifts caused by the mixing
effects. The effective vertex can be constructed from the
Lorentz invariance and the parity and charge-conjugation
symmetries as
\begin{eqnarray}
\Lag_{\gam \pv} &=&
\frac{ g_{\pv} }{ m_{0} } e \tilde{F}^\ext_{\mu \nu} (\partial^{\mu} P) V^{\nu}
\label{eq:L_pv}
\ ,
\end{eqnarray}
where $e>0$ is the unit electric charge, $g_{\pv}$ the
dimensionless phenomenological coupling constant and
$m_0=\frac{1}{2}(m_{\ps}+m_{\V})$ with $m_{\ps}$ and $m_{\V}$
being the vacuum masses of the $\eta_c$ and $J/\psi$, respectively.
We find that effective couplings 
with $F_{\mu\nu}^\ext$, 
such as $F_{\mu\nu}^\ext (\partial^\mu S) V^\nu$ with a scalar field $S$, 
vanish for charmonia at rest in $B$-fields 
due to vanishing components, $\partial^i = 0$ and $F^\ext_{0i} = - F^\ext_{i0} = 0$, 
and also that the static $\eta_c$ is mixed only with the longitudinal$J/\psi$ 
that is polarized in parallel to the external $B$-fields, 
as $\tilde F^\ext_{03} (\partial^0 P) V^3$. 
The coupling constant $g_{\pv}$ can be fitted
to the measured radiative decay width as 
$g_{\pv} =  \sqrt{ 12 \pi e^{-2}  p^{ - 3 }_{f}  m_{0}^{2} \;
\Gamma_{\rm exp}  [ \Jp \to \gamma \etac ] \; } = 2.095 $
with $p_f= (m_\V^2 - m_\ps^2) / (2 m_\V)$
being the magnitude of the center-of-mass momentum in the final state.

Introducing a constant $B$-field in Eq.~(\ref{eq:L_pv}), we solve
the two-state problem for the $\eta_c$ and the longitudinal
$J/\psi$ using the classical Euler-Lagrange equation of
$\mathcal{L}_{\text{eff}} =
\mathcal{L}_{\text{kin}}+\mathcal{L}_{\gam \pv}$. We obtain the
physical mass eigenstates in the presence of the mixing effects 
\begin{equation}
m_{\Jp,\etac}^{2}=\frac{1}{2} \bigg( M_+^2+\frac{\gamma^2}{m_{0}^{2}}
\pm \sqrt{M_-^4+\frac{2\gamma^2 M_+^2}{m_0^2}+ \frac{\gamma^4}{m_0^4}} \bigg), \label{eq:EFT}
\end{equation}
where $M_+^2=m_\ps^2+m_\V^2, M_-^2=m_\V^2-m_\ps^2$ and
$\gamma=g_{\pv} e B$. Expanding Eq.~\eqref{eq:EFT} up to the
second order in $\gamma$ and the leading order in 
$\frac{1}{2}(m_\V-m_\ps)/m_{0} $ 
, we find
\beq 
m_{\Jp,\etac}^{2} &=& 
m_{\V,\ps}^{2} \pm \frac{\gamma^2}{M_-^2}  
, 
\label{eq:Jpsi_2nd}
\eeq
with eigenvectors given by
\beq
| \etac )_{\scriptscriptstyle B} \ \, &=&  \bigg(1-\frac{1}{2} \frac{\gamma^2}{ M_-^4}  \bigg)
| P ) - i \frac{\gamma}{M_-^2}
| V )   , \nonumber \\
| \Jp )_{\scriptscriptstyle B} &=& -i \frac{\gamma}{M_-^2}
| P ) +  \bigg(1-\frac{1}{2} \frac{\gamma^2}{M_-^4} \bigg)
| V ) .
\label{eq:wave}
\eeq
These results show a decrease and an increase in the masses of the $\eta_c$
and the longitudinal $J/\psi$, respectively. Such a level repulsion
has also been found
on the basis of a potential-model approach \cite{AS}.
However, further mass shifts could be caused by $B$-fields acting
on the charmed meson loops such as a $D\bar{D}$ loop and/or
interactions among charmonia and two photons ($B$-fields) as
higher-order corrections to the effective Lagrangian 
(\ref{eq:L_pv}).


To examine the effects of external $B$-fields on the charmonia using a
nonperturbative QCD formalism, we turn to the QCDSR. 
We consider the current correlators in external $B$-fields for
the PS current $J^\ps = i \bar c \gamma^5 c$ and the vector
current $J^\V_\mu = \bar c \gamma_\mu c$ defined by
\begin{eqnarray}
\Pi ^J (q) = i \int \!\! d^4\!x \, e^{iq\cdot x} \langle 0 \vert T[ J(x) J(0) ] \vert 0 \rangle
\label{eq:corr}
\ \ ,
\end{eqnarray}
where superscripts $J=P$ and $V$ denote the PS and the vector currents, respectively.
We investigate a spin-projected scalar correlator for the longitudinal $J/\psi$,
$\tilde \Pi^\V = (\epsilon^\mu \Pi^{\V}_{\mu\nu} \epsilon^\nu) / q^2$,
specified by a polarization vector $\epsilon^\mu =  (0,0,0,1) $
in a $B$-field 
oriented in the third spatial direction.
The PS correlator is normalized as
$\tilde \Pi^\ps = \Pi^\ps/q^2$.
We will construct the sum rules for $\tilde\Pi^J(q^2)$.

The first step involves calculating the OPE in the presence of an
external $B$-field. The OPE for the HQ systems is based on the
expansion in the deep Euclidean region $Q^2 = -q^2 \gg 0$, where $\vert
\langle Op \rangle \vert \ll 4m^2 + Q^2  $, with the left-hand
side being the typical scale of the vacuum and/or the external
field. Thus, as long as the $B$-field satisfies the similar
condition $\vert e \bm B \vert \ll 4m^2 + Q^2 $,
which is valid for 
a region $\vert e \bm B \vert \alt 10 m_\pi^2$ expected 
up to Large Hadron Collider energies \cite{Bestimates}, we can include the effect as
an additional OPE term to the conventional terms in the ordinary
vacuum \cite{RRY81,RRYrev} 
\begin{eqnarray}
\tilde\Pi_\ope (Q^2) = \tilde\Pi^\vac_\ope (Q^2) +
\tilde\Pi_\ope^\ext (Q^2)
. 
\label{eq:Pve0} \ \
\end{eqnarray}
The correlator $ \tilde\Pi_\ope^\ext $ can be precisely evaluated to the second order of $eB$
by utilizing the corresponding coefficients for the dimension-4 gluon
condensates \cite{RRY81,RRYrev,KKLW} with an appropriate correction of
the color matrix factor $t^a$, i.e.,
$ \tr[t^a t^a] \langle G^a_{\mu\nu} G^a_{\alpha\beta} \rangle
\rightarrow \tr[\id_{\rm color}] \langle F_{\mu\nu} F_{\alpha\beta} \rangle $.
We show the Borel-transformed Wilson coefficients 
for static charmonia with $q^i = 0$ 
in Eqs.~(\ref{eq:Cp}) and (\ref{eq:Cv}), 
and full accounts of the calculation in a subsequent paper \cite{paper}.
In the extremely strong field limit $\vert e \bB \vert \gg 4m^2 + Q^2  $,
one has to go beyond the ordinary perturbation theory 
and perform resummation over the all-order dimension operators 
as recently investigated by one of the present authors 
in the vector channel \cite{HI}.

Another possible effect of external $B$-fields on the correlator
is a modification of the gluon condensate $\langle G^{a \mu\nu}
G^a_{\mu\nu} \rangle $. Recently, both a lattice QCD simulation
\cite{BMW_GG} and an analytic study \cite{Ozaki} pointed out that
the gluon condensate increases with an increasing $B$-field at zero temperature 
in analogy to the growth of the quark condensate in magnetic fields
known as {\it magnetic catalysis} \cite{GMS, BMW_qbarq}.
However, we do not take this into account in the present work,
because this effect should be small 
without direct couplings between gluons and external $B$-fields,
as estimated to be a change less than 10 \% \cite{BMW_GG}.



The current correlator \eqref{eq:corr} is connected to the
physical spectral density $\rho(s) = \text{Im}\tilde\Pi(s)/\pi$ in
the deep Euclidean region $Q^2$ through the dispersion relation
\begin{equation}
 \tilde\Pi^J(Q^2) = \int ds \frac{\rho(s)}{s+Q^2} +
 (\text{subtraction}).
  \label{eq:disp}
\end{equation}
In the QCDSR, the phenomenological side for $\rho(s)$ is
often modeled by 
the ground-state pole and the continuum.
The sum rule is known to be insensitive to the structure of
$\rho(s)$ in the high-energy perturbative regime after the Borel
transformation of the dispersion relation. However, if there is a
mixing with a state close to the ground state, it should be
carefully included in the phenomenological side. Below, we show
how to accomplish this in the case of the $\eta_c$; the same
calculation can be straightforwardly applied to the longitudinal $J/\psi$.

We start by considering the low energy states that interpolate the
currents in the correlation function \eqref{eq:corr} with the
PS current. Since the $J/\psi$ mixes into the PS correlator in the
second order of $eB$,
we have 
\begin{eqnarray}
\tilde\Pi^{\ps}_\ph(q^2) &=& \frac{| \langle 0|J^5| \etac
\rangle|^2   }{q^2-m_{\etac}^2} +\frac{| \langle 0|J^5| \Jp
\rangle|^2   }{q^2-m_{\Jp}^2}, \label{eq:ph}
\end{eqnarray}
where the matrix element is calculated in the presence of the
external $B$-field as follows. Let us first look at the residue
of the second term. The current can either couple directly to the
$\Jp$ or first couple to the $\etac$ which will be subsequently
converted to the $\Jp$ through the hadronic coupling given in Eq.~\eqref{eq:L_pv}. 
These can be written as
\begin{eqnarray}
| \langle 0|J^5| \Jp \rangle|^2= f_\dir +  \frac{ f_0 | \langle P| \Jp \rangle|^2   }{(q^2-m_{\ps}^2)^2}.
\label{eq:fp}
\end{eqnarray}
with $f_\dir = \vert \langle V  \vert  J^5(q) \vert 0 \rangle \vert^2$
and $f_0 = \vert \langle P \vert J^5(q) \vert 0 \rangle \vert^2$.
The effective vertex \eqref{eq:L_pv} leads to $| \langle P|
\Jp \rangle|^2 =\gamma^2$. Using the Bethe-Salpeter amplitudes
\cite{BSamp} with the Coulombic wave function of the S-wave
quarkonia, we compute the direct-coupling through a triangle
diagram \cite{paper} as $f_\dir =a_0^4 Q_c^2/64(eB)^2 f_0$
with the electromagnetic charge of a charm quark $Q_c = 2/3$ and
the Bohr radius $a_0 = 0.811$ GeV$^{-1}$, chosen to fit the
root-mean-square radius of the $\Jp$ obtained from the Cornell 
potential model. After inserting Eq.~(\ref{eq:fp}) to Eq.~\eqref{eq:ph}, 
we find that the second term in Eq.~\eqref{eq:ph} can now be decomposed as
\begin{align}
\label{eq:exp}
&\hspace{-0.25cm}
\frac{f_0 \gamma^2}{ (q^2-m_{\ps}^2)^2 (q^2-m_{\V}^2) }
\nonumber\\
& \ \ =
\frac{f_0 \gamma^2}{M_-^4 }
\left[\frac{1}{q^2 - m_{\V}^2} - \frac{1}{q^2 - m_{\ps}^2}
- \frac{ M_-^2 }{ (q^2-m_{\ps}^2)^2 }  \right]
,
\end{align}
where the $\Jp$ mass was replaced by the vacuum mass $m_\V$ within
the second-order correction in $eB$ to the correlator \eqref{eq:ph}. 
The strength of the vector single pole is found to be
much larger than the direct-coupling strength, $f_\dir / (f_0
\gamma^2/M_-^4) \sim  0.0003 $, so that one can safely neglect the
contributions of the direct couplings, including cross diagrams
in which a $J/\psi$ converted from a $\eta_c$ is directly coupled
to the PS current.

One should note that the phenomenological side discussed above can be
obtained by first converting the current to the psedoscalar meson with
the strength $f_0$ and then using the second-order perturbation theory
shown in Eqs.~(\ref{eq:Jpsi_2nd}) and (\ref{eq:wave}),
in which the correlator is given by
\begin{eqnarray}
\Pi_{\rm 2nd}(q^2) =
f_0  \left[ \
\frac{\vert ( P \vert \etac )_{\scriptscriptstyle B} \vert^2}{q^2-m_\etac^2}
+ \frac{\vert ( P \vert \Jp )_{\scriptscriptstyle B} \vert^2}{q^2-m_\Jp^2}
\ \right]
\label{eq:2nd}
.
\end{eqnarray}
All three of the terms in  Eq.~\eqref{eq:exp} are reproduced by
expanding the rhs in Eq.~(\ref{eq:2nd}) up to the second order
in $eB$. Interpretation of the terms in Eq.~\eqref{eq:exp} are as
follows. The first term corresponds to production of an on-shell
$J/\psi$ from the PS current via an off-shell $\eta_c$. The second
term with a negative sign is needed to preserve the normalization,
because the coupling of $\eta_c$ to the current must be reduced to
balance the occurrence of the coupling to $J/\psi$. This is
confirmed in Eq.~(\ref{eq:2nd}), where these two terms come from
overlaps between the properly normalized unperturbated and
perturbated states obtained as $\vert ( P \vert \etac
)_{\scriptscriptstyle B} \vert^2 \sim 1 - (\gamma/M_-^2)^2$
and $\vert ( P \vert \Jp )_{\scriptscriptstyle B}
\vert^2 \sim (\gamma /M_-^2)^2$.
The third term has a double pole on the $\eta_c$ mass with a
factor $M_-^2$, corresponding to a virtual transition to $J/\psi$
state between on-shell $\eta_c$ states, which is nothing but the
origin of the mass shift due to the mixing effect. In
Eq.~(\ref{eq:2nd}), this term comes from an expansion with respect
to the mass correction shown in Eq.~(\ref{eq:Jpsi_2nd}). Clearly, if one
includes this mixing term in the phenomenological spectral function, 
its effect is subtracted out from the total mass shift obtained from the QCDSR, 
and thus can be separated from the residual effects of $B$-fields, 
{\it not described in the hadronic level}.

Now we evaluate the mass spectra of $\eta_c$ and $J/\psi$ using the standard Borel transformation method. 
With a transformation parameter $M^2$ called the Borel mass, the
dispersion relation \eqref{eq:disp} is transformed to
\begin{equation}
 \mathcal{M}^J(M^2) = \int_{0}^{\infty} ds \, e^{-s/M^2} \text{Im}\tilde\Pi^J(s) ,
\label{eq:Boreltransform}
\end{equation}
so that the sum rule can be expressed as ($\nu = 4m_c^2/M^2$)
\begin{equation}
 \mathcal{M}_{\text{OPE}}(\nu) =
  \mathcal{M}_{\text{ph}}^{\text{pole}}(\nu) + \mathcal{M}^{\text{cont}}(\nu) +
  \mathcal{M}^{\text{ext}}_{\text{ph}}(\nu)
.
\label{eq:Boreldispersion}
\end{equation}
A transform of the OPE side (\ref{eq:Pve0}) is then
obtained as
\begin{align}
 \mathcal{M}_{\text{OPE}}(\nu) = &\pi e^{-\nu}A(\nu)[1+\alpha_s(\nu)a(\nu) \nonumber\\
 &+ b(\nu)(\phi_b + \phi_b^{\text{ext}})+ c^{\text{ext}}(\nu)\phi_c^{\text{ext}} ]
.
\label{eq:OPE}
\end{align}
While explicit forms of the coefficients $A(\nu)$, $a(\nu) $ and $b(\nu)$
are given in Refs.~\cite{Bertl,ML10},
the Lorentz-breaking part $c^{\text{ext}}(\nu)$ is obtained for static charmonia 
$(q^i = 0)$ to be
\begin{align}
 c^{\ps,\text{ext}}(\nu)& = \frac{4}{3}b(\nu) - \frac{16}{3}\nu \frac{\Whittaker{-\frac12}{\frac12}{\nu}}{\Whittaker{\frac12}{\frac32}{\nu}}
 \label{eq:Cp},
\\
 c^{\V,\text{ext}}(\nu)&=
 \frac{2\nu}{3\Whittaker{\frac{1}{2}}{\frac{5}{2}}{\nu}}\left[
 6\Whittaker{\frac12}{\frac52}{\nu} \nonumber\right.\\
 &\hspace{0.5cm}
 \left. +6\Whittaker{-\frac12}{\frac52}{\nu}-\Whittaker{-\frac32}{\frac52}{\nu}
 \right]
 \label{eq:Cv}
 ,
\end{align}
with $G(a,b,\nu)$ being the Whittaker function. 
Operator expectation values
$\phi_b^\text{ext}$ and $\phi_c^{\text{ext}}$ account for
magnitudes of the external $B$-fields, and are defined by 
$\phi_b^{\text{ext}} = \frac{4}{3}\frac{Q_c^2}{16m_c^4}(eB)^2$
and
$\phi_c^{\text{ext}}= -\frac{Q_c^2}{16m_c^4}(eB)^2$.

As for the phenomenological side on the rhs of
Eq.~\eqref{eq:Boreldispersion},
$\mathcal{M}_{\text{ph}}^{\text{pole}}$ and
$\mathcal{M}^{\text{cont}}$ have the same form as in the
conventional QCDSR analyses. While
$\mathcal{M}_{\text{ph}}^{\text{pole}}$ corresponds to the
transform of the first term in Eq.~\eqref{eq:ph} given by
$\mathcal{M}_{\text{ph}}^{\text{pole}} = f_0
e^{-m^2_{\eta_c}/M^2}$, $\mathcal{M}^{\text{cont}}$ stands for a
perturbative continuum contribution
$\theta(s-s_0)\text{Im}\tilde\Pi(s)$ up to $\mathcal{O}(\alpha_s)$
with $s_0$ being the effective threshold parameter. The
$B$-dependent part $\mathcal{M}_{\text{ph}}^{\text{ext}}$
considered above is, by inserting the correlator \eqref{eq:ph} 
into Eq.~\eqref{eq:Boreltransform}, obtained as
\begin{eqnarray}
\M_{\ph}^{\text{ext},\etac} (M^2) &=& f_0 (eB)^2 \left[ \
Q_c^2 \frac{a_0^4}{64} e^{- \frac{m_\V^2}{M^2}}
\right.\label{eq:Mph-etac}
\\
&&\hspace{-0.7cm}
\left.
+ \frac{ g_\pv^2}{M_-^4}
 \left( \, e^{- \frac{m_{\V}^2}{M^2}} - e^{- \frac{m_\ps^2}{M^2}}
+ \frac{ M_{-}^2}{ M^2} e^{- \frac{m_\ps^2}{M^2}} \, \right)
\ \right]
\nonumber
.
\end{eqnarray}

The corresponding formula for $J/\psi$ can be obtained by
interchanging $m_{\ps}$ and $m_{\V}$. Following from a sign flip
in $M_-^2$, we find that the double-pole contribution in the
vector channel has the opposite sign to that of the last term in
Eq.~\eqref{eq:Mph-etac}. Inserting these results into the
Borel-transformed dispersion relation (\ref{eq:Boreldispersion}), 
the mass of the lowest-lying pole can
be evaluated from an equation,
\begin{equation}
m_{\eta_c}^2(M^2) = -\frac{\partial}{\partial(1/M^2)}\ln[\mathcal{M}_{\text{OPE}}-\mathcal{M}^{\text{cont}}_{\text{ph}}-\mathcal{M}_{\text{ph}}^{\text{ext}}]
.
\end{equation}
We examine a stability of $m^2$ with respect to the
$M^2$-dependence, called the Borel curve, within a range of $M^2$
which satisfies two competing conditions, that is, less than 30\%
contribution from the dimension-4 opeators to the OPE and more
than 70\% lowest-pole dominance in the dispersion integral \eqref{eq:Boreltransform}, 
specifying the Borel window. The
effective threshold parameter $s_0$ is so tuned to make the Borel
curve the least sensitive to $M^2$.
Finally, we average the value of the mass over the Borel window 
and calculate the variance to estimate a systematic error.
See Ref.~\cite{ML12} for the details of the systematic framework.

 \begin{figure}[!t]
 \vspace{-0.2cm}
  \includegraphics[width=3.375in]{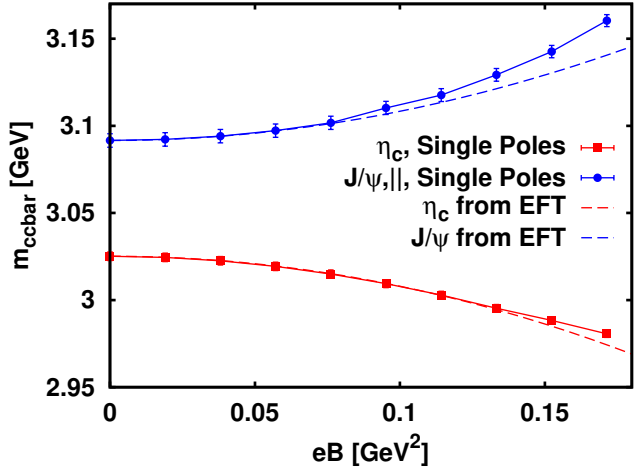}
    \vspace{-0.6cm}
  \caption{Mass of the charmonium states from the QCD sum rules (closed
  symbols with solid lines) and the effective Langrangian
  \eqref{eq:Jpsi_2nd} 
  (dashed lines) as functions of $eB$.}\label{fig:mass-B}
  \vspace{-0.2cm}
 \end{figure}

With $\alpha_s(8m_c^2)=0.24$, $m_c(p^2=-2m_c^2)=1.26$ GeV and
$\langle \frac{\alpha_s}{\pi}G^2 \rangle = (0.35 \ \text{GeV})^4$, the
vacuum mass of $J/\psi$ and $\eta_c$ are found to be 3.092 GeV and
3.025 GeV, respectively. 
To compare results from the QCDSR with those from the effective
Lagrangian \eqref{eq:L_pv}, 
we insert these vacuum masses into $m_{\ps,\V}$ in Eq.~\eqref{eq:Jpsi_2nd}.
The effective coupling 
$g_{\pv}$ obtained above is used in both approaches. 
Figure \ref{fig:mass-B} displays the results from the QCDSR with the
phenomenological side shown on the rhs in
Eq.~\eqref{eq:Boreldispersion}, but without including the
double-pole term responsible for the mixing effect in
$\mathcal{M}_{\text{ph}}^{\text{ext}}$ [see Eq.~\eqref{eq:Mph-etac}].
Remarkably, one finds perfect agreement
between the two approaches in $eB < 0.1$GeV$^2$,
followed by a slight deviation as $eB$ is further increased.
The agreement indicates that the 
$B$-dependent terms in Eq.~\eqref{eq:Mph-etac}
are essential ingredients to obtain physically meaningful results in QCDSR,
where the level repulsion is understood as a
consequence of the different signs of the single-pole terms in Eq.~\eqref{eq:Mph-etac},
$e^{-m^2_{\V}/M^2} < e^{-m^2_{\ps}/M^2}$,
owing to the vacuum mass difference.

 \begin{figure}[!t]
 \vspace{-0.2cm}
  \includegraphics[width=3.375in]{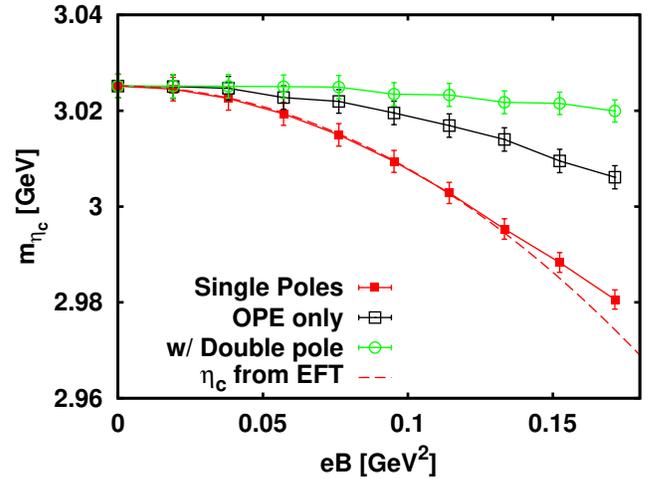}
  \vspace{-0.6cm}
  \caption{Mass of $\eta_c$ from the QCD sum rule with different
  implementations of the phenomenological side. }
  \label{fig:mass_phen}
 \end{figure}

In order to understand the role of each term in $\mathcal{M}_{\text{ph}}^{\text{ext}}$,
we perform the QCDSR analyses in two cases
employing the phenomenological sides without $\mathcal{M}_{\text{ph}}^{\text{ext}}$
and with all the terms of $\mathcal{M}_{\text{ph}}^{\text{ext}}$ including the double-pole term.
In Fig.~\ref{fig:mass_phen}, we show results for the $\eta_c$ in these two cases with open symbols.
Without any $B$-induced poles, 
one obtains the open squares (``OPE only''),
which show somewhat heavier mass than
the final results obtained by including the single poles
(filled squares in Figs.~\ref{fig:mass-B} and \ref{fig:mass_phen}). 
Since the conventional one-peak 
spectral ansatz cannot account for
the occurrence of the $J/\psi$ pole induced by $B$-fields,
the resultant $\eta_c$ mass is an average of $\eta_c$ and $J/\psi$, 
giving the artificially heavier $\eta_c$ mass.
On the other hand, if one includes all the terms of $\mathcal{M}_{\text{ph}}^{\text{ext}}$,
the $\eta_c$ mass becomes almost constant, despite the fact that
the magnetic field contribution is included in the OPE. This means
that the double-pole term on the phenomenological side in Eq.
\eqref{eq:Mph-etac} almost exclusively accounts for 
the $B$-dependence on the OPE
side. The residual mass shift, albeit tiny for the $\eta_c$, is 
an effect that cannot be explained by the mixing effect.

In conclusion, we have discussed effects of strong magnetic fields on
the mass spectra of $\etac$ and $\Jp$ 
with an elaborate treatment of the mixing effects 
on the phenomenological side in the QCDSR. 
We found
that the mass shifts are dominated by the
level repulsion coming from the mixing effect 
in precise agreement with those from the effective Lagrangian
approach. While the residual mass shift is found to be small for
the charmonia, our analysis indicates that, to obtain the correct
results, one has to take into account effects of the magnetic fields 
on the phenomenological side as well as the OPE side. 
Therefore, a similar approach should be adopted when
investigating light mesons by QCDSR or even any other systems
involving the spectral density by means of the correlation
functions in constant magnetic fields.


\textit{Acknowledgements}.---
This work was supported by the Korean Research Foundation 
under Grants No.~KRF-2011-0020333 and No. KRF-2011-0030621. 
K.M. is supported by HIC for FAIR, partially by the Polish Science Foundation (NCN) 
under Maestro Grant No. 2013/10/A/ST2/00106 and by the Grant-in-Aid 
for Scientific Research on Innovative Areas from MEXT (No. 24105008)
The research of K.H. is supported by JSPS Grants-in-Aid No.~25287066. 
S.C. was supported by the Korean Ministry of Education through the BK21 PLUS program.
Three of the authors (K.H., K.M. and S.O.) thank Yukawa Institute for Theoretical Physics, Kyoto University,
where a part of this work was discussed 
during the YIPQS international workshop ``{\it New Frontiers in QCD 2013}.''



\end{document}